# Analogous response of temperate terrestrial exoplanets and Earth's climate dynamics to greenhouse gas supplement

Assaf Hochman[1*], Thaddeus D. Komacek[2], and Paolo De Luca[3]

1. Fredy and Nadine Hermann Institute of Earth Sciences, The Hebrew University of Jerusalem, Jerusalem, Israel.

2. The University of Maryland, Department of Astronomy, College Park, USA.

3. Barcelona Supercomputing Center, Barcelona, Spain.

*Corresponding author: Assaf Hochman

**Abstract**

Humanity is close to characterizing the atmospheres of rocky exoplanets due to the advent of JWST. These astronomical observations motivate us to understand exoplanetary atmospheres to constrain habitability. We study the influence greenhouse gas supplement has on the atmosphere of TRAPPIST-1e, an Earth-like exoplanet, and Earth itself by analyzing ExoCAM and CMIP6 model simulations. We find an analogous relationship between $CO_2$ supplement and amplified warming at non-irradiated regions (night side and polar) - such spatial heterogeneity results in significant global circulation changes. A dynamical systems framework provides additional insight into the vertical dynamics of the atmospheres. Indeed, we demonstrate that adding $CO_2$ increases temporal stability near the surface and decreases stability at low pressures. Although Earth and TRAPPIST-1e take entirely different climate states, they share the relative response between climate dynamics and greenhouse gas supplements.

**Keywords:** Habitability, James Webb Space Telescope, Dynamical systems theory, Chaos, Climate change, Arctic amplification, Polar amplification

**Introduction**

A fundamental understanding of Earth's climate dynamics is critical for assessing the impacts of climate change. The recent detections of a panoply of potentially habitable planets (e.g., Proxima Centauri b[1], TRAPPIST-1e,f,g[2], Wolf 1069b[3], LP 890-9c[4]) have opened a window to study the climates of a broad range of planets that may be Earth-like but have much smaller and cooler host stars. Importantly, due to their close-in orbits, these planets are expected to be tidally locked to their host star, with a dayside in perpetual sunlight and a nightside in constant darkness. This day-to-night irradiation contrast will greatly change the climate dynamics of rocky planets orbiting late-type M dwarf stars relative to those orbiting Sun-like stars[5, 6, 7]. As a result, these nearby exoplanets provide an opportunity to improve our theoretical understanding of planetary climate through a combination of first-principles modeling and observational constraints with current and future state-of-the-art facilities (e.g., James Webb Space Telescope - JWST, Extremely Large Telescopes - ELT, Large Interferometer For Exoplanets - LIFE, and the Habitable Worlds Observatory - HWO).

TRAPPIST-1e is a prime habitable exoplanet with a 6.1-day orbital period around an ultracool M dwarf star that has been studied in detail with previous 3D GCM simulations. Note that these GCM simulations implicitly assume an atmospheric composition that would allow for habitable surface conditions, but it is possible that the UV emission during the bright pre-main-sequence phase of M dwarf stars causes atmospheric loss along with water photolysis and resulting escape to space[8, 9, 10]. These GCMs have predicted that TRAPPIST-1e will likely have a temperate climate for a wide range of possible background carbon dioxide supplements [e.g., 11, 12, 13]. Additionally, the TRAPPIST-1 Habitable Atmosphere Intercomparison protocol has compared the predictions of four GCMs (ExoCAM, LMD-Generic/PCM, ROCKE-3D, UK Met Office Unified Model) and found broad agreement, with ExoCAM having the most humid and cloudy atmosphere in the protocol experiments[11]. A range of model predictions has also found that key atmospheric constituents, including carbon dioxide and methane, may be detectable on TRAPPIST-1e with the JWST Near Infrared Spectrograph (NIRSpec) instrument, enabling constraints on its habitability[13, 14, 15]. Previous work has further demonstrated that TRAPPIST-1e has greater climate variability than Earth, which necessitates understanding the fundamental climate drivers of tidally-locked planets[16, 17, 18].

On Earth, The Intergovernmental Panel on Climate Change (IPCC)[19] has used the results from multiple phases of the Coupled Model Intercomparison Project (CMIP)[20] in its assessments of the state of the global climate and its projections for the future. CMIP3, CMIP5, and CMIP6 are the three phases of this initiative, and they have played a critical role in advancing our understanding of the Earth's climate system[21, 22, 23]. These fully coupled atmospheric-oceanic models simulate various climate system components, including wind and temperature, and are used to analyze past and future Earth climate changes. These changes can significantly impact multiple aspects of the climate system, including precipitation, ocean currents, and ecosystems. The CMIP results have shown that increasing concentrations of greenhouse gasses are driving significant changes in global wind and temperature patterns[19, 24, 25]. In the latest CMIP6 simulations, the influence of changes in greenhouse gasses and adaptation policies on the climate system has been assessed using scenarios termed Shared Socioeconomic Pathways (SSP)[26].

In the CMIP3 and CMIP5 experiments, models have projected that the global temperature and wind patterns will change in response to increasing greenhouse gas concentrations, leading to alterations in the intensity and location of jet streams, a slowdown of the summer circulation in the northern mid-latitudes and an increase in the global moisture budget[27, 28, 29]. In the more recent CMIP6 project, the models have improved in their ability to simulate these changes and provide more detailed projections of temperature and wind changes at regional scales[30, 31, 32]. The CMIP6 models also project that many areas will experience amplified warming, particularly in the high latitudes and the Arctic[33]. These changes may influence baroclinicity and the location of storm tracks and Rossby-wave structure[34, 35].

Dynamical systems theory provides a robust mathematical framework for understanding the behavior of complex systems over time, including the climate of Earth and exoplanets[18]. Earth's and exoplanet's atmospheric dynamics are chaotic, i.e., non-linear and extremely sensitive to their initial conditions[36, 37]. It is, thus, impossible to characterize the dynamics of atmospheres analytically. This is because non-linear systems cannot be broken into parts and solved separately[38]. This problem can be overcome using a geometric point of view, portraying the dynamics of atmospheres in phase space[39]. In this study, we describe the atmospheres of Earth and TRAPPIST-1e by relying on recent developments in dynamical systems theory[40]. These advancements have allowed us to describe atmospheric patterns in terms of local dimension ($d$) and inverse persistence ($\theta$), which provide insights into the atmosphere's 'stability' (here, in the context of the atmosphere's ability to change between states). The local dimension ($d$) can be interpreted as a proxy for the number of degrees of freedom active around the atmospheric state, i.e., the number of options the atmospheric state can evolve into or from. Inverse persistence ($\theta$), instead, measures the mean residence time of the system around the state. In the case of discrete-time data, the value of $\theta$ is influenced by the time interval between data points, and its value ranges between zero and one. However, if a trajectory quickly moves away from the vicinity of state $\xi$, the value of $\theta(\xi)$ is closer to one. Accordingly, a highly persistent (low $\theta$) low dimensional (low $d$) state will be more stable than a low-persistence (high $\theta$), high-dimensional (high $d$) one (Fig. S1)[40]. Further details of how $d$ and $\theta$ are computed are provided in the Methods section. The dynamical systems framework has been applied to various climate datasets [e.g., 41, 42, 43]. Indeed, Hochman et al.[18] have demonstrated that this framework can be used at the planetary scale for both Earth and TRAPPIST-1e.

The primary aim of this study is to determine how greenhouse gas supplement influences atmospheric dynamics and climate variability. To do so, we test the impact of carbon dioxide partial pressure ($pCO_2$) on the climate dynamics of both Earth-like planets orbiting Sun-like stars and Earth-sized planets orbiting late-type M dwarf stars that are in close-in orbits, resulting in spin-synchronization. Specifically, we conduct a suite of ExoCAM General Circulation Model (GCM) simulations of the prime habitable exoplanet candidate TRAPPIST-1e and Earth-analog exoplanets orbiting Sun-like stars, both with varying $pCO_2$. We determine the differences between the climates of TRAPPIST-1e and Earth with a wide range of carbon dioxide supplements and draw similarities between their climatology. We then provide a dynamical systems analysis of the climate dynamics of our TRAPPIST-1e and Earth-like ExoCAM, as well as CMIP6 simulations, to study how varying the abundance of greenhouse gasses affects the vertical structure and variability of their circulation.

Results

**Comparing temperature and circulation changes of TRAPPIST-1e and Earth-like simulations with varying $pCO_2$ and vertical levels**

We first analyze the mean climatology of the Earth-like simulations in terms of temperature and circulation (Fig. 1 a-d). For comparison, the climatology of ERA5 reanalysis (see data) and CMIP6 historical simulations and projections are shown in Fig. 2. The low $pCO_2$ scenario ($10^{-2}$ bar) near the surface captures the Earth's temperature patterns and global circulation (cf. Fig. 1 d to Fig. 2 a). Indeed, the simulation shows higher temperatures at the equatorial regions with respect to the poles. In addition, the north and southeast trade winds at equatorial regions and the westerlies at the mid-latitudes are adequately replicated, considering that the Earth-like simulation is assumed to be an aqua-planet with no seasonality (cf. Fig. 1 d to Fig. 2 a). Inspecting the upper level circulation, we find that the Polar and sub-tropical jet streams have average wind speeds of ~50 m s$^{-1}$, comparable to present-day Earth (cf. Fig. 1 b to Fig. 2 b).

Next, we analyze the same as above but for TRAPPIST-1e (Fig. 1 e-h). Since TRAPPIST-1e is assumed to be a tidally locked aqua-planet, we find that low ($10^{-2}$ bar) and high (1 bar) $pCO_2$ scenarios are characterized by an "eyeball" climate state with temperatures peaking near the substellar point and decreasing toward the nightside[44]. This eyeball-like region near the surface spans most latitudes from 45W to 45E longitudes in both $pCO_2$ scenarios (Fig. 1 g, h).

After inspecting the mean climatology of the Earth-like and TRAPPIST-1e simulations, we analyze the influence $pCO_2$ has on their atmospheric circulation. Figure 3 displays the mean temperature and wind speed differences between high and low $pCO_2$ scenarios for Earth-like and TRAPPIST-1e simulations at upper and near-surface levels. To determine the statistical significance of the disparity in the composite maps, we conduct a Wilcoxon Rank-Sum test[45]. This test evaluates the medians of two datasets assuming the null hypothesis that they are identical. Thus, a statistically significant p-value at the 5% level indicates that the medians are distinct. We find a decrease (increase) in wind speed near the surface at mid-latitude (polar) regions with increasing $pCO_2$ of the Earth-like simulation (Fig. 3 c). Apparent changes in wind direction are also shown. On the contrary, the upper levels show mainly increases in wind speeds at both mid-latitudes and polar regions (Fig. 3 a). We relate this to a meridional decrease (increase) in near-surface (upper level) temperature gradients due to polar amplification (Fig. 3 a, c)[46]. Interestingly, the same phenomenon is observed in the TRAPPIST-1e simulation even though it is assumed to be tidally locked, with no received irradiation on the nightside (Fig. 3 b, d). Indeed, we find amplified warming in the non-irradiated near-surface regions of TRAPPIST-1e (Fig. 3 d). The upper level shows a mirror image with an Eastern shift of the eye-like irradiated region due to increased $pCO_2$ (Fig. 3 b and Fig. 1 e, f). Since this is not the case near the surface (cf. Fig. 1 g and h), some regions on TRAPPIST-1e may be influenced by increased baroclinicity due to vertical differences in the shift of the planetary-scale wave pattern with increasing $pCO_2$.

Accordingly, the regions subject to a decrease in temperature gradients also show reductions in wind speed and vice versa, which are comparable to the Earth-like simulation values. Finally, $pCO_2$ variation significantly influences the day-to-night temperature contrast and resulting wave structure of TRAPPIST-1e. Increasing greenhouse forcing is expected to decrease temperature contrasts on tidally locked planets due to an increase in longwave radiation re-radiated to the nightside of the planet in combination with increased heat transport from dry static energy fluxes[47]. Both the low and high $pCO_2$ cases have Rossby gyres at low pressures on the nightside due to the day-night temperature contrast inducing a high-amplitude Matsuno-Gill pattern (cf. Fig. 1 g and h; Fig. 3 d)[48]. The Matsuno-Gill pattern has a similar structure between the low and high $pCO_2$ cases, however, the root-mean-square (RMS) horizontal wind speed at the upper level is 55.4 m s$^{-1}$ in the case with high $pCO_2$ and 13.0 m s$^{-1}$ in the case with low $pCO_2$. This relative increase in wind speeds at the upper level in the high pCO$_2$ case is concomitant with an increase in the horizontal temperature contrasts at the upper level.

Next, we analyze the global mean temperature and wind speed vertical profiles of the Earth-like and TRAPPIST-1e simulations with varying pCO$_2$ (Fig. 4 a-b). We find a greater near-surface temperature sensitivity of TRAPPIST-1e (+70 K) to pCO$_2$ compared to the Earth-like (+50 K) simulations (Fig. 4 a)[18]. This is, however, not the case at the upper levels, where the Earth-analog case is more sensitive to an increase in pCO$_2$ (Fig. 4 a). A more complex picture arises from inspecting the influence an increase in pCO$_2$ has on the vertical structure of average wind speed (Fig. 4 b). The Earth-like and TRAPPIST-1e simulations provide evidence for a global average decrease (increase) in wind speed near the surface (upper level) with increasing pCO$_2$. The average changes are more prominent in the Earth-like (~+50 m s$^{-1}$) simulation than in TRAPPIST-1e (~+35 m s$^{-1}$; Fig. 4 b).

The above findings suggest an analogous response of both Earth and TRAPPIST-1e to an increase in pCO$_2$, particularly concerning changes in circulation, i.e., wind speed, direction, and horizontal temperature contrasts. We, therefore, turn to compare our idealized simulations to CMIP6 simulations. We first compare the CMIP6 multi-model ensemble median (MMEM) vertical profiles to ERA5 reanalysis. The CMIP6 models capture the vertical structure of temperature and wind speed (cf. Fig. 4 c, d with e, f). Then, we show that the CMIP6 MMEM displays an increase (decrease) in temperature near the surface (upper level) with the change in the shared socioeconomic pathway (SSP; Fig. 4 c and Fig. 2 c-f)[19]. The opposite holds when inspecting changes in the vertical profile of wind speed, i.e., a slight decrease (increase) in wind speed near the surface (upper level; Fig. 4 d and Fig. 2 c-f). Though the magnitude of the changes in the CMIP6 models is relatively small compared to our idealized simulations, since we consider an order of magnitude pCO$_2$ supplement (see Data and Methods), the direction of the change in the near-surface and upper level for both temperature and wind speed is the same in all three cases.

**Comparing vertical dynamical changes of TRAPPIST-1e and the Earth-like simulations with varying pCO$_2$**

We examine the changes in the vertical dynamics of TRAPPIST-1e, the Earth analog, and Earth itself due to changes in greenhouse gas supplement using a dynamical systems perspective (Fig. 5). Thus, we compute dynamical systems metrics for the 2-dimensional average temperature (T) and average wind speed (WS) variables at various vertical levels (see Methods). We find that with increasing pCO$_2$, the atmospheric persistence of TRAPPIST-1e and the Earth analog will increase (decrease in $\theta$) near the surface and decrease in the upper atmosphere (increase in $\theta$; Fig. 5 a, b). When considering the change of local dimension ($d$), as a response to an increase in pCO$_2$, we find that for the Earth analog, there is a decrease in $d$ near the surface and an increase at upper levels (Fig. 5 c, d), i.e., lower atmospheric changeability near the surface, whereas, higher at upper levels. TRAPPIST-1e, however, shows an increase in $d$ at all vertical levels (Fig. 5 c, d).

Comparing the CMIP6 model simulations to ERA5 reanalysis, we find that the vertical dynamics in CMIP6 are very well captured. This allows us to examine the influence pCO$_2$ has on the vertical dynamics and then compare it to the Earth analog and TRAPPIST-1e. The $d$ and $\theta$ metrics in the CMIP6 simulations show a decrease in $d$ and $\theta$ near the surface and an increase at upper levels, similar to the Earth analog, though with a relatively smaller amplitude (cf. Fig. 5 e-h with a-d). Again, we relate this to the relatively small changes in greenhouse supplement in the SSP scenarios compared to the order of magnitude changes we consider in our idealized simulations (see Data and Methods). The same is shown when comparing the response of the $\theta$ metric between TRAPPIST-1e and CMIP6 simulations (cf. Fig. 5 e-f with a-b). The response of $d$, however, is similar in CMIP6 and TRAPPIST-1e only at upper levels, both displaying an increase in $d$ (cf. Fig. 5 g-h with c-d).

In summary, we find deviations in the time series dynamics of the global circulation with increasing pCO$_2$ on TRAPPIST-1e, the Earth analog, and Earth itself.

**Summary and conclusions**

Observations of habitability indicators and biosignatures in the atmospheres of exoplanets provide an avenue to constrain the prevalence of habitable conditions in our galaxy[49, 50, 51]. One of the critical aspects of exoplanet habitability is understanding their atmospheres compared to Earth[52]. Recently, the precision and spectral coverage of exoplanet atmospheric spectroscopy has significantly increased due to the commissioning of the James Webb Space Telescope (JWST). The telescope has enabled broad-wavelength infrared spectroscopy of close-in exoplanets[53, 54, 55], and promises to deliver detailed spectroscopic observations of rocky exoplanet atmospheres in transmission[56]. In this work, we applied a suite of idealized ExoCAM model simulations of TRAPPIST-1e and Earth to understand the influence greenhouse gas supplement has on their temperature, atmospheric circulation, and dynamics. We further analyzed an ensemble of CMIP6 model simulations and ERA5 reanalysis for comparison.

Our key findings and conclusions are as follows:

1. Although TRAPPIST-1e and Earth take entirely different mean climate states, they share the same response of amplified warming at non-irradiated regions sensitive to greenhouse gas supplements.

2. Amplified warming at non-irradiated regions on TRAPPIST-1e and Earth changes the temperature gradients and significantly varies global circulation. Changes in temperature gradients may influence baroclinicity in some areas on both terrestrial bodies[57].

3. A dynamical systems framework provides additional insights into the atmosphere's response to greenhouse gas supplements. Indeed, increasing greenhouse gas abundance increases persistence near the surface[18] and decreases it at upper levels. This finding may have significant implications for understanding how Earth's atmosphere will respond as it warms. Specifically, such changes may influence the evolution of the climate's mean state and extremes.

4. Changes with orders of magnitude in $pCO_2$ on Earth lead to different climate states and dynamics, especially as a function of height. Such changes may imply how Earth's climate evolved and will evolve as the inner edge of the Solar System's habitable zone shifts as the Sun brightens.

5. The vertical temperature, wind structure, and persistence strongly depend on the $pCO_2$ supplement. As a result, we anticipate that observations of both tidally locked rocky exoplanets orbiting M dwarf stars with JWST and ELTs, as well as observations of Earth-sized planets at ~1AU separations orbiting Sun-like stars with HWO and LIFE, may be able to constrain the amplitude of the resulting climate variability. Constraints on climate variability of temperate rocky exoplanets would open a window to constraining their climate state and habitability.

In conclusion, with the commissioning of JWST and extremely large ground-based telescopes, there is an expectation that astronomers characterize rocky exoplanets in detail. Exoplanetary atmospheres are incredibly diverse and can vary significantly compared to Earth's atmosphere. This diversity has opened up new research avenues and led to a greater understanding of the atmospheres on other planets and on Earth. In this work, we have compared Earth to one specific nearby exoplanet target, TRAPPIST-1e. Therefore, our results are limited in scope, and further work is required to determine whether increasing greenhouse gas concentration ubiquitously amplifies the warming at non-irradiated regions of rocky planets with Earth-like atmospheres. We envisage that our analysis framework can be applied to a broad range of exoplanetary atmospheres, including model inter-comparison protocols[11], and may provide additional insights into how Earth's atmosphere has and will evolve.

**Data**

*ERA5 reanalysis*

We use the European Center for Medium-Range Weather Forecast (ECMWF) ERA5 reanalysis[58] from 1981-2010. ERA5 reanalysis is a state-of-the-art global atmospheric reanalysis dataset developed by the European Centre for Medium-Range Weather Forecasts (ECMWF). Such dataset combines observational data with a numerical weather model to create a consistent and comprehensive record of past weather conditions. We extract temperature (T) and (WS) over 37 vertical pressure levels, from 1000 to 10 hPa. The ERA5 reanalysis data are at 0.25° × 0.25° grid spacing (or longitude-latitude 1440 × 720), and we re-grid them to 5° × 3.9° (or longitude-latitude 72 × 46) to facilitate comparison with the ExoCAM simulations (see Methods).

*Earth climate simulations*

We use a multimodel ensemble (MME) of nine Coupled Model Intercomparison Project Phase 6 (CMIP6) historical simulations (1981-2010) and future (2071-2100) high-emission Shared Socioeconomic Pathway (SSP 5-8.5)[23, 26]. We use the first ensemble member to extract T and WS over eight vertical pressure levels, from 1000 to 10 hPa. We compute the WS from meridional (V) and zonal (U) winds. We then re-grid the datasets to 5° × 3.9° (or longitude-latitude 72 × 46) to make it comparable with the ExoCAM simulations. The CMIP6 models and ensemble members used are listed in Table S1.

**Methods**

*ExoCAM simulations*

To simulate the atmospheric dynamics of TRAPPIST-1e and an Earth-analog exoplanet, we use the ExoCAM GCM[59], publicly available at https://github.com/storyofthewolf/ExoCAM along with its radiative transfer package, ExoRT (https://github.com/storyofthewolf/ExoRT). ExoCAM is an established exoplanet GCM that modifies the Community Atmosphere Model (CAM version 4)[59] and extends the GCM to include varying planetary and orbital properties and a novel correlated-k radiative transfer scheme that allows for simulations of hotter climates than modern Earth. ExoCAM has been used for a broad range of studies of rocky exoplanet climates [e.g., 6, 7, 18], and has been compared to other rocky exoplanets in GCMs via the TRAPPIST Habitable Atmosphere Intercomparison (THAI) protocol[11, 61, 62].

In this work, we conducted ExoCAM simulations of TRAPPIST-1e and an Earth-analog exoplanet with a similar setup over an equivalent range of greenhouse gas supplements. The initial conditions for these simulations are identical to those in Hochman et al.18. They use the same assumed planetary parameters. However, here each simulation is extended for 30 years to study the vertical dependence of climate variability. Specifically, we conducted simulations for two distinct values of carbon dioxide partial pressure (pCO$_2$), 10$^{-2}$ bars (low) and 1 bar (high). We chose these values to cover temperate and hot climates for TRAPPIST-1e 11. Our simulations of TRAPPIST-1e assume a stellar spectrum of a late-type M dwarf with an effective temperature of 2600 K [63], while our Earth-analog models use a Solar-like host star spectrum.

Our TRAPPIST-1e and Earth analog simulations use the same fundamental assumptions while consistently varying the stellar incident flux, stellar spectrum, surface gravity, and rotation rate to correspond to each planet. Both models assume an aqua-planet surface with a slab ocean of 50 m depth, providing abundant water vapor set via the Clausius-Clapeyron relationship. All models include 1 bar of background N$_2$, implying that the total atmospheric pressure varies between our models with varying pCO$_2$. All simulations assume zero obliquity and zero eccentricity (i.e., no seasons) to facilitate inter-comparison between the tidally locked assumption for TRAPPIST-1e and the fast-rotating Earth-analog. As in Hochman et al.18, all simulations use a horizontal grid spacing of 5° × 3.9° (lon-lat) with 40 vertical levels and a dynamical timestep of 30 minutes.

*Dynamical systems metrics*

We assess the dynamical characteristics of T and WS over the vertical levels of TRAPPIST-1e, Earth, CMIP6, and ERA5 reanalysis datasets with a method grounded in dynamical systems theory. This approach combines Poincaré recurrences with extreme value theory[40, 64], and it has been applied successfully to various climate fields and datasets on Earth [41, 42, 43, 65], and recently also to TRAPPIST-1e18.

This approach provides information about the evolution of a given atmospheric variable (in our case, T and WS) in the atmosphere's phase space. It permits the computation of instantaneous characteristics of chaotic dynamical systems over a given longitude-latitude map (in our case Earth and TRAPPIST-1e). We base the dynamical systems framework on two metrics with a temporal resolution comparable to the atmospheric data (here, we consider daily output): the local inverse persistence ($\theta$) and local dimension ($d$) 40. The metric $\theta$ of a specific atmospheric state would approximate the mean residence time of the trajectory in a small region in phase space. At the same time, $d$ estimates the number of options a given atmospheric state can transit from and to. The lower $\theta$, the more persistent the atmospheric state is, and the lower the $d$, the fewer possibilities the atmospheric state can transition to and from (Fig. S1). Therefore, low $\theta$ and low $d$ values signify

'stable' (in the sense of change between atmospheric states) atmospheric configurations, whereas high $\theta$ and high $d$ represent lower 'stability.' For more detailed information on our dynamical systems approach, we refer the reader to Hochman et al.[18].

## Acknowledgments

PDL was partially funded by the European Union's Horizon Europe Research and Innovation Program under grant agreement 101059659. AH would like to thank the Hebrew University for its technical support. All authors thank Margarida Samso-Cabre for downloading, storing, and formatting the CMIP6 and ERA5 datasets used in the analyses. The authors acknowledge the University of Maryland supercomputing resources (http://hpcc.umd.edu) made available for conducting the ExoCAM climate simulations reported in this paper.

## Author contributions

AH, TDK, and PDL conceived the study. TDK performed the ExoCAM climate model simulations. PDL, TDK and AH prepared and analyzed the data. TDK, PDL and AH prepared the figures. AH, TDK and PDL wrote the manuscript.

## Competing interests

The authors have no competing interests to declare.

## Data and code availability

The datasets used and/or analysed during the current study are available from the corresponding author on reasonable request. The dynamical systems analysis code is freely available at https://ch.mathworks.com/matlabcentral/fileexchange/95768-attractor-local-dimension-and-local-persistence-computation.

12 | Page

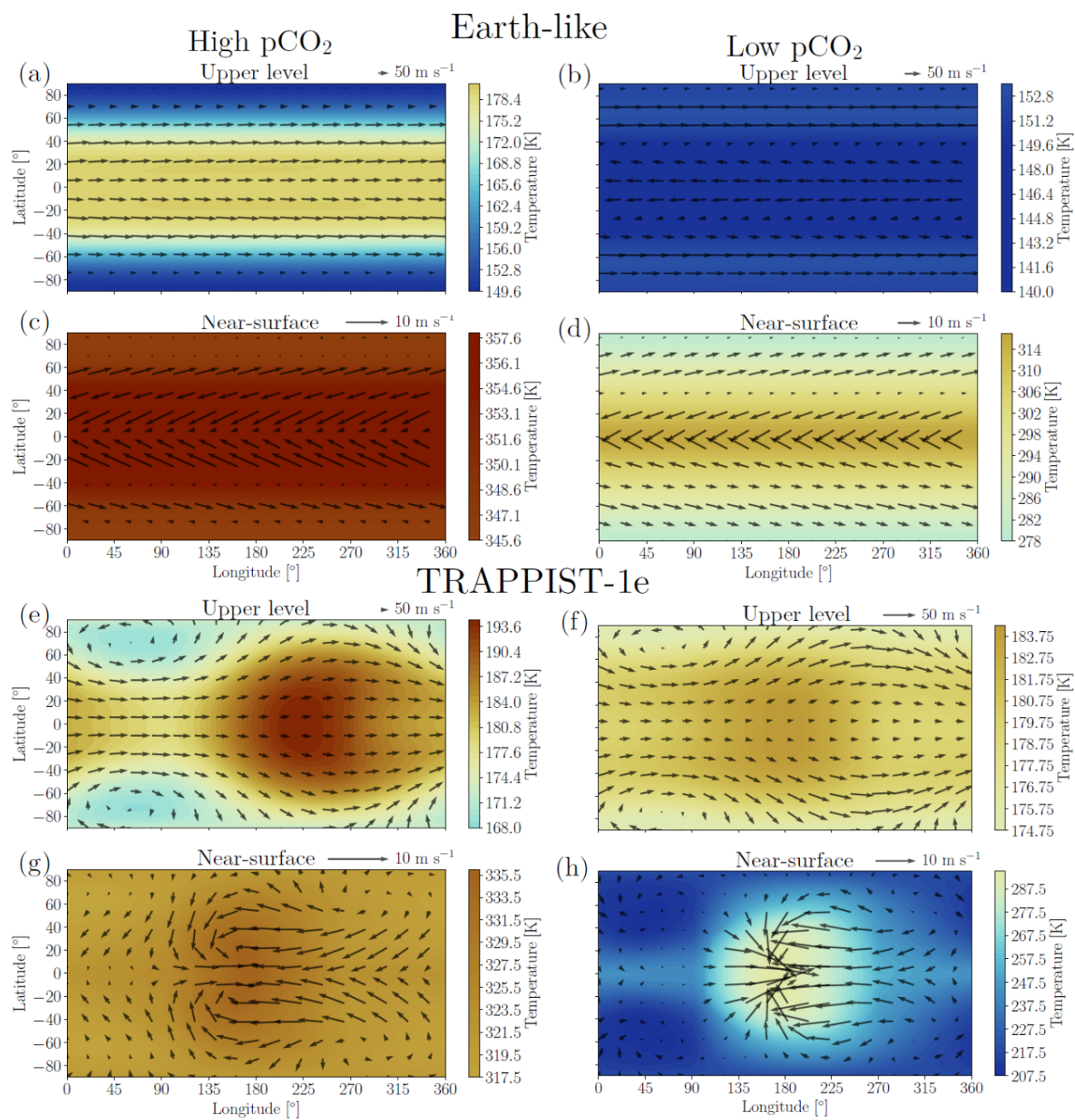

**Figure 1** Climatology of average temperature (K, colors) and wind speed (m s$^{-1}$, quivers) for Earth-like (a - d) and TRAPPIST-1e (e - h) ExoCAM simulations for two CO$_2$ partial pressure (pCO$_2$) scenarios. (a, c, e, g) High pCO$_2$ = 1 bar; (b, d, f, h) Low pCO$_2$ = 10$^{-2}$ bar. Two vertical model levels are presented. (a, b, e, f) upper level at $\sigma = 0.956 \times 10^{-3}$, and (c, d, g, h) near-surface $\sigma = 0.992$. The longitude of the substellar point in the TRAPPIST-1e simulations (e-h) is 180°. Note that maps on the two vertical levels share separate color scales.

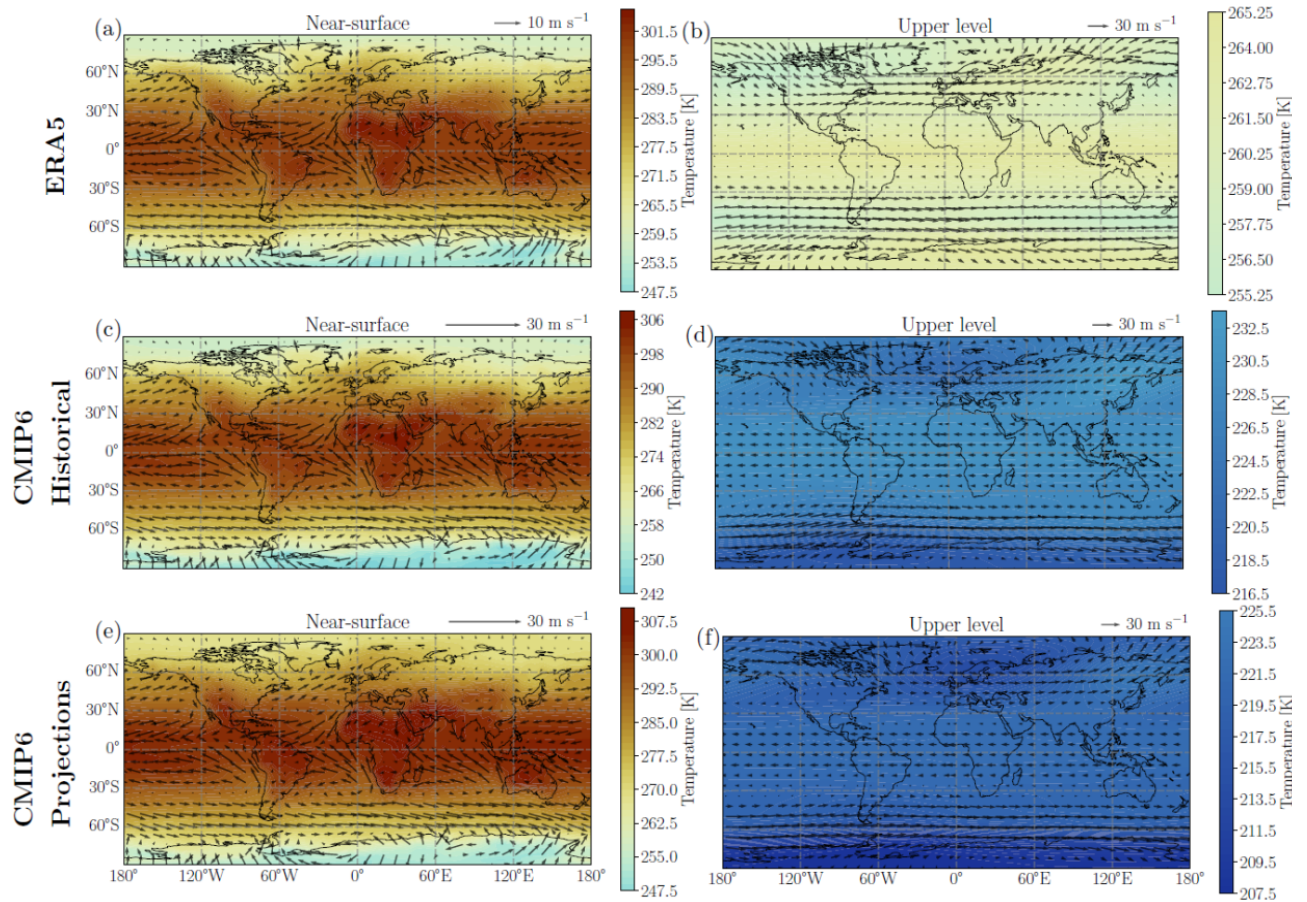

**Figure 2** ERA5 (a, b), CMIP6 historical climatology from 1981-2010 (c, d) and CMIP6 climate projections from 2071-2100 (e, f) for temperature and wind speed at pressure levels of 1 hPa (b, d, f, 'upper level') and 1000 hPa (a, c, e, 'near-surface'). All panels share a color scale but have individual color bars displaying the plotted range of temperatures. (a, b) are means, whereas (c-f) are multi-model ensemble medians (MMEM).

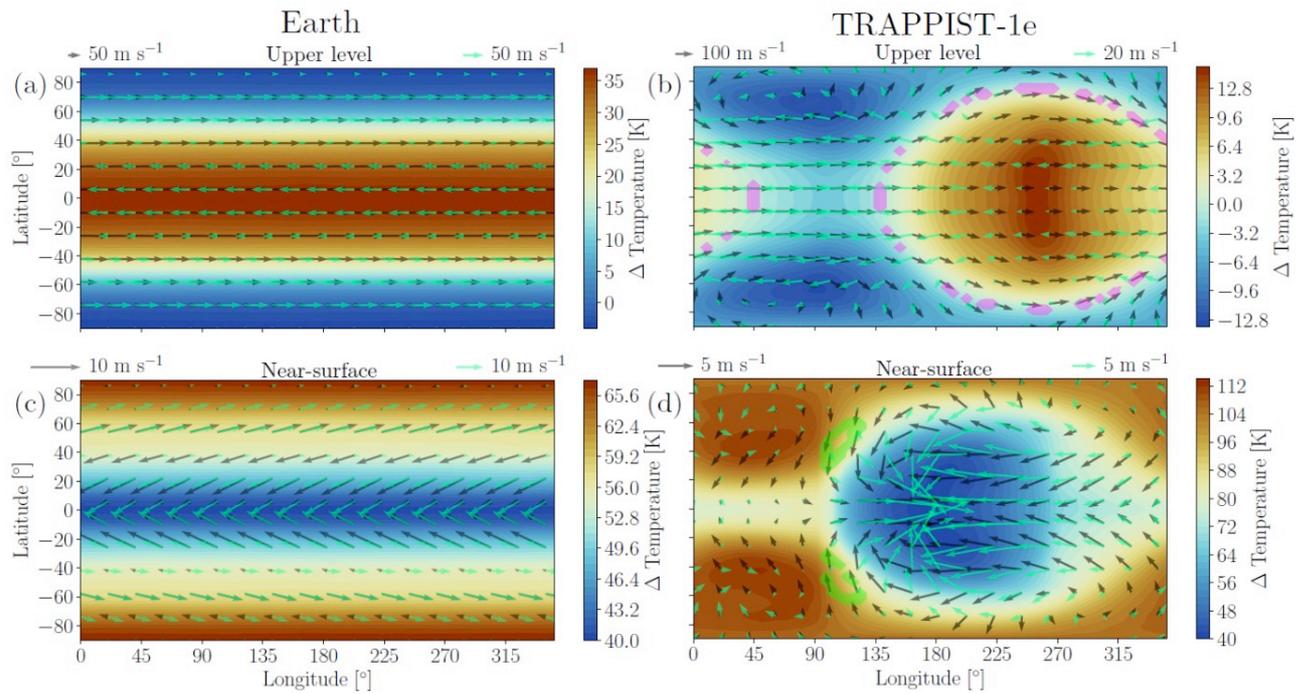

**Figure 3** Median temperature (K) differences between high and low $pCO_2$ scenarios and median wind speeds (m s$^{-1}$) for each scenario for Earth-like (a, c) and TRAPPIST-1e (b, d) ExoCAM simulations at (a, b) upper and (c, d) near-surface levels. Black and green quivers represent the high and low $pCO_2$ scenarios, respectively. Using the Wilcoxon Rank-Sum test, lime and magenta-filled contours for wind speed and temperature represent areas not statistically significant at the 5% level, respectively.

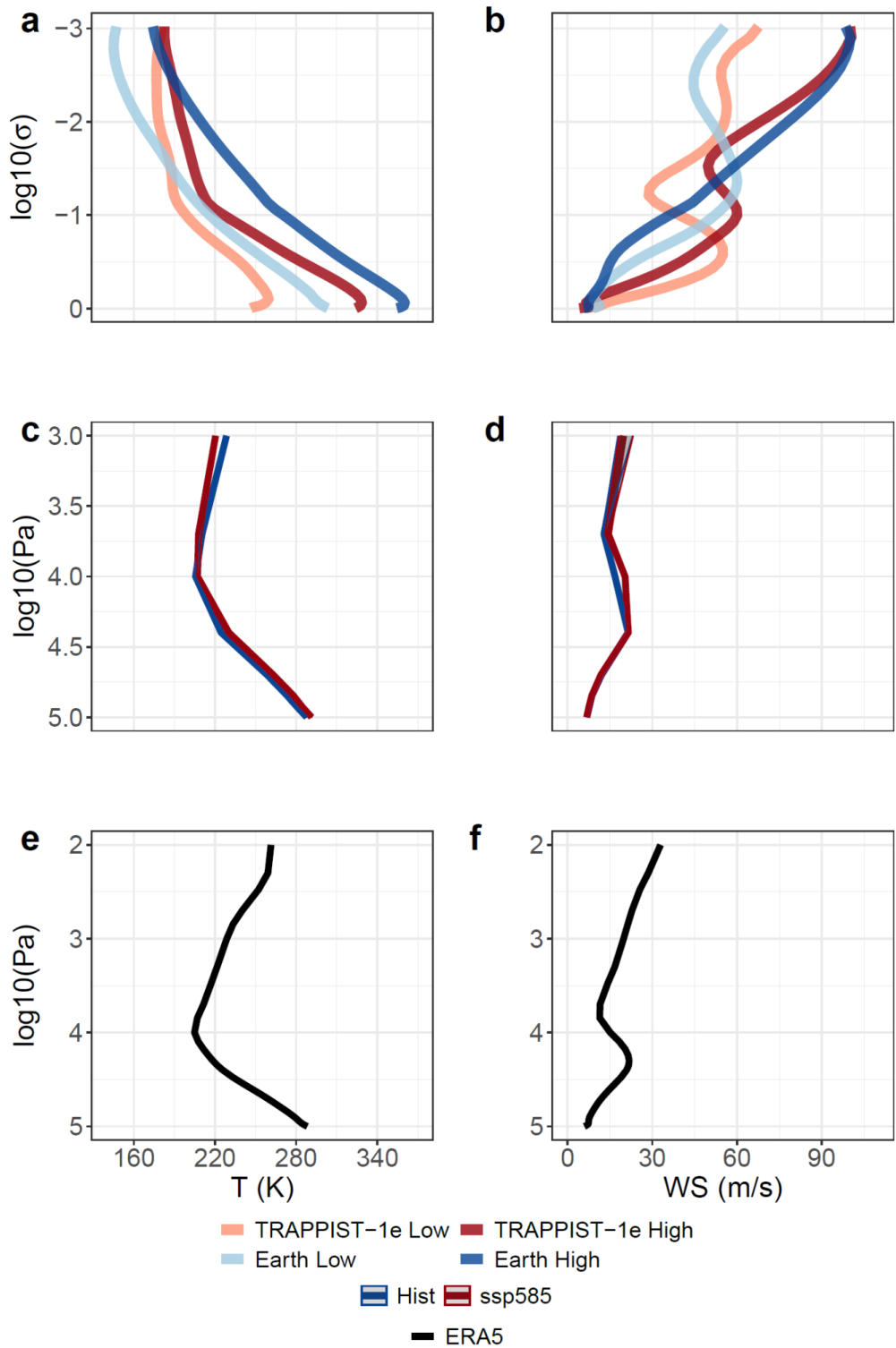

**Figure 4** Global temperature (T in K) and wind speed (WS in m s$^{-1}$) vertical profiles. (a, b) TRAPPIST-1e and Earth-like ExoCAM simulation averages under high and low pCO$_2$ scenarios. (c, d) CMIP6 MMEM for the historical (1981-2010) and SSP5-8.5 (2071-2100) periods. (e, f) ERA5 reanalysis average for the historical period (1981-2010). (a, b, e, f) show averages, whereas (c, d) the multi-model ensemble medians, with shaded bands representing the interquartile range (25th and 75th percentiles) of the MME.

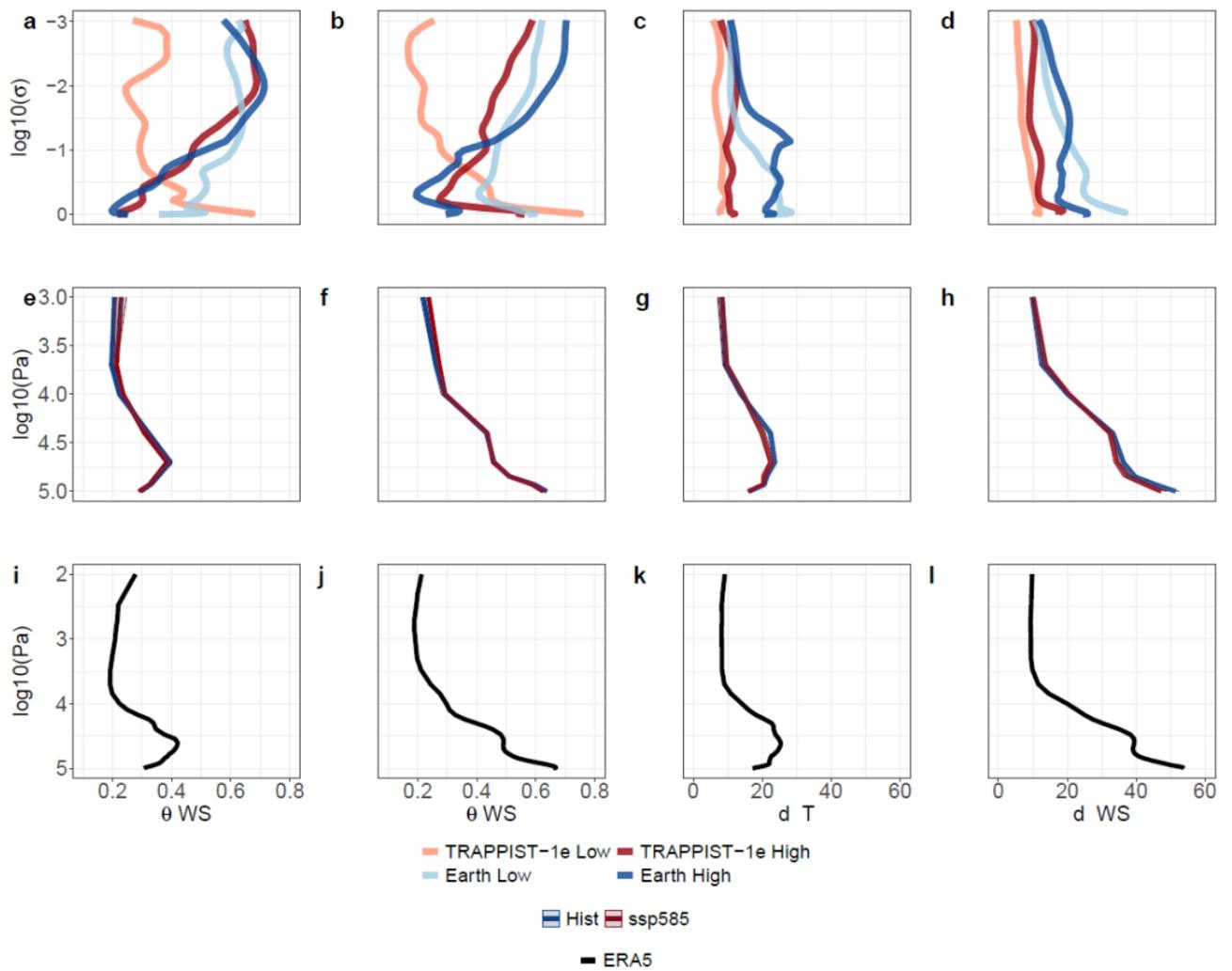

**Figure 5** Global dynamical systems metrics vertical profiles for temperature (T) and wind speed (WS). (a - d) TRAPPIST-1e and Earth-like ExoCAM simulation averages under high and low $pCO_2$ scenarios. (e - h) CMIP6 multi-model ensemble medians for the historical (1981-2010) and SSP5-8.5 (2071-2100) periods. (i - l) Average of ERA5 reanalysis for the historical period (1981-2010). Dynamical systems metrics are local inverse persistence ($\theta$) and local dimension ($d$; see Methods).